\documentstyle{article}
\title{
\begin{flushright}
{\bf\normalsize   COLO-HEP-286}\\
\end{flushright}
\bf The XY Model Coupled to Two-Dimensional Quantum Gravity
}

\author{ {\it C.F. Baillie} \\
         Physics Dept. \\
         University of Colorado\\
         Boulder, CO 80309, USA\\
	 \\
         and \\
	 \\
         {\it D.A. Johnston}\\
         Dept. of Mathematics\\
         Heriot-Watt University\\
         Riccarton\\
         Edinburgh, EH14 4AS, Scotland}

\begin{document}
  \maketitle
                      {\Large
                      \begin{abstract}
%
We perform Monte Carlo simulations using the Wolff
cluster algorithm of the XY model on both fixed and dynamical
phi-cubed graphs (i.e. without and with coupling to two-dimensional
quantum gravity).
We compare the numerical results with the theoretical expectation
that the phase transition remains of KT type when the XY model
is coupled to gravity. We also examine whether the universality we
discovered in our earlier work on various Potts models with
the same value of the central charge, $c$, carries over to the XY
model, which has $c=1$.
\\
\\
Submitted to Phys Lett B.
%
                        \end{abstract} }
%
  \thispagestyle{empty}
%
%
  \newpage
%
                  \pagenumbering{arabic}

\section{Introduction}

In the past few years the work of KPZ in the lightcone gauge Liouville theory
\cite{1}
and DDK in the conformal gauge \cite{2} has allowed the calculation
of the critical exponents for $c \le 1$ conformal models coupled
to 2d quantum gravity. These conformal field theory results are backed up by
the
exact solution of the Ising model, with $c=1/2$, using matrix model methods in
\cite{3}, which gives identical
exponents to those calculated in \cite{1},\cite{2}. Numerical simulations of
the Ising model on
both dynamical triangulations and dynamical phi-cubed graphs \cite{4}, which we
would expect to discretize
the continuum theory coupled to gravity, also give good agreement with these
analytical predictions.
We have recently simulated the $Q=3,4$ state Potts models on dynamical
phi-cubed graphs, with $c=4/5, 1$
respectively, and again found good agreement with the conformal field theory
predictions \cite{5}
(in these cases there are, as yet, no exact solutions available).

All of the analytical approaches to 2d quantum gravity, whether by continuum
Liouville
theory or matrix models, break down at $c=1$. We can see this quite clearly in
the KPZ/DDK results
for the ``dressed'' conformal weight $\Delta$ of an operator with weight
$\Delta_0$ before
coupling to gravity
\begin{equation}
\Delta - \Delta_0 = - {\alpha^2 \over 2} \Delta ( \Delta - 1),
\label{e0}
\end{equation}
where $\alpha$ is
\begin{equation}
\alpha = - { 1 \over 2 \sqrt{3} } ( \sqrt{ 25 -c } - \sqrt{ 1 -c} ).
\label{e01}
\end{equation}
We thus get nonsensical complex weights for $c>1$. Similarly, $d=0$ multimatrix
models and
the multicritical points of the $d=0$ single matrix model appear to give access
only to $c<1$.
There have been various speculations regarding the nature of the ``barrier'' at
$c=1$
\cite{6} and suggestions on extending the validity of the Liouville theory to
$c>1$ \cite{7},
although it is interesting to note that both numerical \cite{8}
and analytical \cite{9} work with multiple Potts models (with $c>1$, at least
naively)
have failed to see any clear signal of pathological behavior.

In spite of this caveat it is clear that $c=1$ represents a particularly
interesting case. In the
matrix model approach we can
construct a $c=1$ model by considering matrices in $d=1$
which gives
the partition function
\begin{equation}
Z = \int D \Phi \exp \left[ - \beta \int dt Tr \left( {1 \over 2} \dot \Phi^2 +
U ( \Phi ) \right) \right]
\label{e02}
\end{equation}
where $\Phi$ is an $N \times N$ hermitian matrix. This model was solved in the
$N \rightarrow \infty$ limit
in \cite{10} and shown to be equivalent to $c=1$ matter coupled to 2d quantum
gravity in \cite{11}.
More recently the
double scaling limit of the model was taken in \cite{12}. As the model's
continuum limit
is an uncompactified boson it is natural to try and extend the analysis
to the case of a compactified boson, which gives us the XY model coupled to 2d
quantum gravity.

The standard XY model on a regular lattice (i.e. {\it not} coupled to 2d
gravity)
with a partition function of the form
\begin{equation}
Z= \prod_i \int d \theta_i \exp \left( \beta \sum_{<ij>} \cos ( \theta_i
-\theta_j) \right)
\label{e03}
\end{equation}
displays an infinite order ``Kosterlitz-Thouless'' (KT) phase transition
caused by the liberation of vortices \cite{13}.
At this transition the correlation length diverges as
\begin{equation}
\xi = A_\xi \exp \left( { B_\xi \over (T - T_c)^\nu } \right)
\label{e04}
\end{equation}
and the susceptibility as
\begin{equation}
\chi = A_\chi \exp \left( { B_\chi \over (T - T_c)^\nu } \right),
\label{e05}
\end{equation}
where the critical exponent $\nu$ is predicted to be $1/2$ \cite{13}.
$T$ is the temperature defined with $k=1$ so that $\beta = 1/T$.
This KT theory also predicts the correlation function critical exponent
$\eta = 1/4$, where $\eta$ is given by
\begin{equation}
\chi = C \xi^{2-\eta}.
\label{e06}
\end{equation}
In spite of some earlier controversy \cite{14}, apparently due to
difficulties in distinguishing KT fits from second order ones,
more recent numerical simulations support these results \cite{15}.

Returning to the XY model coupled to 2d gravity, the compactified model was
solved in \cite{16}
for a singlet sector in which the angular integrations in the
matrix integral were unimportant. This was identified with the vortex free
sector of the XY model. The adjoint sector was then incorporated into the model
in \cite{17}
and identified with a single vortex/anti-vortex pair. The predictions of
\cite{17}
for the XY model coupled to 2d quantum gravity were, in direct analogy with the
standard XY
model, a KT phase transition at $\beta_c = 2 \pi$. It is interesting to note
that the phase
transition is predicted to remain the same when coupled to gravity, unlike the
case of the Ising and $Q=3,4$
state Potts models where second order phase transitions are weakened to third
order. In this paper
we simulate the XY model on dynamical phi-cubed graphs
\begin{equation}
Z_N =  \sum_{G^{(N)}} \prod_i \int d \theta_i \exp \left( - \beta
\sum_{i,j=1}^N G^{(N)}_{ij}
\cos(\theta_i -\theta_j)   \right)
\label{e2}
\end{equation}
where $G^{(N)}_{ij}$ is the connectivity matrix of the graph, in order to see
how these predictions
tally with ``experiment''.
In \cite{8} we found universal graph properties for various Potts models,
depending only on $c$, so we also measure these here to see how the XY model
compares with the other
$c=1$ models we have investigated.

\section{Fixed Simulations}

As a check of our program we first simulated the XY model on a {\it fixed}
random
lattice coming from
a pure two-dimensional quantum gravity simulation \cite{18}.
These lattices are phi-cubed
graphs with the topology of a sphere, so each point has exactly three
neighbors.
It is currently unclear whether it is possible to take the continuum limit on
a single graph of this form \cite{18a} because of its highly irregular nature,
but we are principally interested here in using the simulation as a test
to compare with our simulation on a dynamical graph so we do not
concern ourselves with this problem. In any case, a simulation
of the Ising model on such a fixed graph gives results that are compatible
with the standard square lattice Onsager exponents \cite{19}.
Our simulations use the Wolff cluster
algorithm \cite{20} to update the spins which is much more effective in
ameliorating
critical slowing down than traditional (local) Monte Carlo methods.
XY models on random triangulations have been simulated in the past
using the Langevin method \cite{21} and the Monte Carlo Renormalization
Group method \cite{22}. The random lattices used in
these previous simulations were obtained by linking up into triangles
points distributed at random in a unit square with periodic boundary
conditions, which gives a much less disordered graph than that coming from
the two-dimensional quantum gravity.

As this is merely a test simulation we used only three small random graphs
with $N=100,500$ and $1000$ points. At $27$ values of $\beta$ between $0.1$
and $3.0$ we ran $10000$ sets of cluster updates
to equilibrate the XY spins and then measured the energy $E$,
specific heat $C$, susceptibility $\chi$ and correlation length $\xi$
over the next $50000$ updates.
(The magnetization $\vec M$ is a vector for the XY model with components
$( \sum_i \cos{\theta_i}, \sum_i \sin{\theta_i} )$ so one usually measures
instead the susceptibility which is defined as its square
$\chi = <\vec M \cdot \vec M> / N$.
The energy, specific heat and correlation length are defined in the usual way.)
Depending on $\beta$ each set of cluster updates consisted of up to $100$
``hits'' of the Wolff algorithm, this is necessary because as $\beta$
decreases (the temperature increases) the size of the cluster built by the
Wolff algorithm decreases therefore more applications of it are required to
update the same number of spins.
To verify that the Wolff algorithm had overcome critical slowing down we
measured the autocorrelation time $\tau$ for the energy,
and found that it increases slightly as $\beta$ increases but does not increase
with $N$, so critical slowing down has essentially been eliminated.
The autocorrelation time in the data is always less than $10$ updates and we
bin this data in bins of size $1000$ for statistical analysis so our error bars
can be
trusted as coming from uncorrelated samples.

In our earlier simulations of single and multiple Potts models coupled to
2d quantum gravity \cite{5,8}, we were able to make use of Binder's cumulant
to determine the critical inverse temperature $\beta_c$ and critical
exponent $\nu$ {\it separately}. Unfortunately this method cannot be
applied to the XY model so we must obtain these from the four-parameter fits
in eqs. \ref{e04},\ref{e05}. As is well known this is rather difficult --
even with very high statistics data on large lattices --
hence the controversy mentioned earlier \cite{14,15}.
Things are made worse by the behaviors of the correlation
length and susceptibility in the XY model:
they both diverge according to eqs. \ref{e04},\ref{e05}
for $T > T_c$ ($\beta < \beta_c$) but are infinite (on an infinite system) for
$T \le T_c$ ($\beta \ge \beta_c$).
Therefore there are no peaks at $T_c$ and on a finite
lattice $\xi$ and $\chi$ just keep increasing as $T$ decreases, with any
jump at $T_c$ rounded away by finite-size effects.
The specific heat does have a peak, which is independent of system size, Fig.
1,
but this occurs at a lower value of $\beta$ than $\beta_c$ so is of no help
in determining $\beta_c$. Hence from our data it is to all intents and purposes
{\it impossible} to determine both $T_c$ ($\beta_c$) and $\nu$. The only way to
proceed is to assume that $\nu = 1/2$ as predicted by KT and attempt to
fit to $T_c$ using the resulting three-parameter fits:
\begin{equation}
\xi = A_\xi \exp \left( { B_\xi \over \sqrt{T - T_c} } \right),\ \ \
\chi = A_\chi \exp \left( { B_\chi \over \sqrt{T - T_c} } \right).
\label{e5}
\end{equation}

To get an idea of what $\beta_c$ (and thus $T_c$)
is approximately we do some finite-size scaling using eq.\ref{e06}.
In finite-size scaling relations one usually has some quantity scaling
as a power of the linear dimension, $L$, of the $d$-dimensional system
with $N=L^d$ points. Here we do not know $d$,
the internal dimensionality of our system, {\it a priori} so must
write instead $N^{1 \over d}$.
The method we shall adopt was used in \cite{22a} to estimate
the exponent $\eta$ knowing $\beta_c$; here we shall use it the other way
around, assuming that $\eta$ takes its KT predicted value of $1/4$.
This is a reasonable thing to do because the XY model has a line of fixed
points for $\beta \ge \beta_c$ along which $\chi$ and $\xi$ diverge
according to eq.\ref{e06}, however, $\eta$ {\it depends on} $\beta$
and in fact the KT analysis predicts $\eta \simeq 1/\beta$,
as $\beta \rightarrow \infty$.
Therefore we only obtain the value $\eta=1/4$ {\it at} $\beta_c$.
The method works as follows. For $\beta \ge \beta_c$ the correlation
length diverges and therefore becomes the system size $N^{1 \over d}$
on our finite lattices. Thus eq.\ref{e06} becomes
\begin{equation}
\chi = C N^{(2-\eta) / d}.
\label{e6}
\end{equation}
Therefore if we plot $[ln(\chi) - ln(C)]/ln(N)]$ versus $\beta$ then
we should obtain ${(2-\eta) / d}$ for $\beta \ge \beta_c$.
At $\beta_c$ where $\eta = 1/4$ and we take $d = 2.6$ from below,
the expected value is $0.673$.
We tune the value for $ln(C)$ until all the points for $\beta \ge \beta_c$
from the different $N$ fall on a universal curve. The resulting plot
(with $ln(C)=1.0$) is shown in Fig. 2, from it we estimate $\beta_c = 2.3(1)$.

We now bite the bullet and attempt to fit to $\xi$ and $\chi$.
For $\xi$ fitting the data between $T=0.5$ and $0.9$ yields
$T_c = 0.36(1)$ with a $\chi^2/dof = 7.9$. At $T=0.5$ the correlation
length has grown to $10$ however which is rather large for the $N=1000$
system. Therefore we restrict the fit to the $7$ points between
$T=0.6$ (where $\xi \approx 6$) and $0.9$ to obtain
$T_c = 0.39(2)$ with a $\chi^2/dof = 0.3$. Taking the difference
as a measure of the error we assume $T_c = 0.39(3)$ so that
$\beta_c = 2.6(2)$. This value is rather high compared with the $2.3$
we obtained from finite-size scaling but, from dynamical results below,
we know that this is a result of having too small a lattice -- on a larger
lattice it would come down to at least (most) $2.4$.
We cannot get nearly as good fits from $\chi$, fitting the same two
temperature ranges yields $T_c = 0.37(2)$ and $0.43(2)$ with enormous
$\chi^2/dof$'s. This fortuitously gives $\beta_c = 2.3(3)$ but again
for larger lattice sizes this value will decrease.
Even for the large-scale simulations of the regular XY model \cite{15}
fits to $\chi$ are almost always worse than fits to $\xi$.
{}From this the best we can conclude is that $\beta_c = 2.3(1)$
for the XY model on a fixed random lattice coming from quantum gravity.

\section{Dynamical Simulations}

For the dynamical simulations, we started from a random phi-cubed graph
coming from pure two-dimensional quantum gravity but updated it during the
simulation using the standard ``flip'' move with the detailed balance
condition that the rings at either ends of the link being flipped have no
links in common. This check eliminates all graphs containing tadpoles or
self-energies. Therefore the Monte Carlo update consists of two parts:
firstly the XY spins are updated by a set of Wolff ``hits'' then $N_{flip}$
links of the graph are picked randomly and flipped.
After testing various values of $N_{flip}$
to ensure that there were enough flips to make the graph dynamical on the
time scale of the XY spin updates we set $N_{flip}=N$ for all the simulations.
We ran on $N=100,500,1000,2000$ and $5000$ at $27$ values of $\beta$
between $0.1$ and $3.0$, doing $10000$ updates for equilibration and
$50000$ for measurement as for the fixed case.
In addition to the standard thermodynamic quantities for the spin model:
$E, C, \chi$ and $\xi$ we measured some properties of the graph:
acceptance rates for flips, distribution of ring lengths and
internal fractal dimension $d$.
We must measure $d$ because we do not know {\it a priori} the internal
dimensionality of our system.
In fact, numerical simulations for pure 2d quantum gravity \cite{23} and for
2d quantum gravity
coupled to Potts models \cite{5,8} yield $d \approx 2.7$;
and an analytical calculation predicts that $d$ lies between $2$ and $3$
for pure 2d quantum gravity \cite{24}.

We measured the autocorrelation time $\tau$ for the energy
for $N=1000, 2000$ and $5000$ and found that
it increases slightly as $\beta$ increases and
with $N$ near the phase transition, though never exceeding $9$.
At the critical point $\beta_c$, where the correlation length becomes the
size of the system, we can fit
\begin{equation}
\tau \simeq N^{z \over d},
\end{equation}
to extract an estimate for the dynamical critical exponent $z$. If we take our
best estimate of $\beta_c$ from below and assume that $d = 2.6$ then we obtain
$z = 0.8(1)$.
This is larger than what we found for Potts models \cite{5,8},
but still a lot less than traditional (local) Monte Carlo methods which
have $z \approx 2$.

As for the fixed simulations we have the same problems in fitting $T_c$
from the correlation length and susceptibility.
We restrict both fits to $8$ points between $T=0.55$ and $0.9$, so
that $\xi < 6$, and obtain the values for $T_c$ listed in columns
2 and 3 of Table 1. We see that, for $\xi$ at least, the $T_c$ values
increase slightly as $N$ increases. Taking the difference between the results
on
$N=2000$ and $N=5000$ as an indication of the error, we thus obtain
$\beta_c = 2.4(2)$ from $\xi$ and
$\beta_c = 2.0(2)$ from $\chi$. The fits for $\xi$ yield
$\chi^2/dof$'s of around $1$, whereas those for $\chi$ again are huge.
Thus we take $2.4(2)$ as the best value for $\beta_c$.
This value is confirmed by our finite-size scaling plot of
$[ln(\chi) - ln(C)]/ln(N)]$ versus $\beta$ in Fig. 3 which yields
$\beta_c = 2.4(1)$.
Finding a $\beta_c$ for the dynamical case that is larger than the fixed case
satisfies naive expectations, since this means that it is more difficult to
``freeze'' the dynamical system (so we must go to a lower temperature).

\begin{center}
\begin{tabular}{|c|c|c|c|} \hline
 $N$ & $T_c$ from $\xi$ & $T_c$ from $\chi$ & $d$ \\[.05in]
\hline
1000 & 0.38(1)          & 0.50(6)           & 2.13(2) \\[.05in]
2000 & 0.39(1)          & 0.51(6)           & 2.25(2) \\[.05in]
5000 & 0.41(1)          & 0.51(5)           & 2.42(1) \\[.05in]
\hline
\end{tabular}
\end{center}
\vspace{.1in}
\centerline{Table 1: Fitted values of $T_c$ and $d$.}
\bigskip

Again the specific heat does have a peak
occurring at a lower value of $\beta$ than $\beta_c$.
However, it now appears that the peak is growing with system size
(which would be the behavior expected from a second order phase transition),
as we can see in Fig. 4.
We think that this is an artifact of not running long enough to obtain
enough bins in order to estimate the specific heat accurately from
the fluctuation in the energy:
\begin{equation}
C = {\beta^2 \over N} (<E^2> - <E>^2).
\end{equation}
Evidence for this is that by using the other way to calculate $C$ --
numerically
differentiating the energy  -- we obtain curves,
which we also plot in Fig. 4, that look exactly
like those from the fixed simulations, where both
methods for obtaining $C$ agree.

Turning now to the graph properties.
We measure the acceptance rate for the flip move
to confirm that our graphs are really dynamical.
The flip can be forbidden either from the
graph constraints coming from the detailed balance condition or from
the energy change of the spin model, so we can decompose the flip acceptance
rate into two parts:
AL -- the fraction of randomly selected links which can be flipped
satisfying the graph constraints; and
AF -- the fraction of links satisfying the graph constraints which are
actually flipped, i.e. pass the Metropolis test using the XY model energy
change.  AF and AL are shown for $N=2000$ in Fig. 5.
We see that they both have dips at some $\beta < \beta_c$.

The distribution of ring lengths in the graph is
the discrete equivalent of the distribution of local Gaussian curvatures
in the continuum.
The minimum possible ring length is 3.
If we also plot the probability (fraction) of rings of length three (PR3) in
Fig. 5
to compare with AF and AL we see that it has a peak very close
to the dip in AL. This is reasonable since both PR3 and AL
depend only on the graph, whereas AF depends on the spin model.
Interestingly the peak in PR3 and dip in AL occur close to the peak
in the specific heat, at a value of $\beta$ lower than $\beta_c$.
In Fig. 6 we compare PR3 with what we obtained for single Potts models
for $N=2000$
\cite{5} in order to see if the universality we
discovered carries over to the XY model (note that we have had
to scale $\beta$ for the XY model by a factor of two to correct for
differing definitions in the actions). If it did we would expect
PR3 for the XY model to look like that for the $Q=4$ Potts model as
they both have central charge $1$. It is rather obvious that PR3 for the XY
model is significantly different, in fact its maximum value of $0.2185$
is rather close to that of the Ising ($Q=2$ Potts) model ($0.2179$).
However its value at the phase transition $0.2160$ lies between those
for the $Q=2$ and $Q=3$ Potts models ($0.2151$ and $0.2167$ respectively),
and a long way from that for the $Q=4$ model ($0.2190$),
so must conclude that there is no universality between the Potts and XY
models coupled to quantum gravity.

Finally we measured the internal fractal dimension $d$ of the dynamical graphs.
We use the most naive definition of distance (the fewest links between two
points
)
so we are considering the ``mathematical geometry'' rather than the ``physical
geometry'' in the terminology of \cite{23} \footnote{It now appears
that the two are, in fact, identical \cite{27}.}.
We measured the internal fractal dimension at $\beta_c$
for all the simulations, obtaining the values in the last column of Table 1
and then extrapolated to $N=\infty$.
There is however some ambiguity in this extrapolation: for Potts models
\cite{5,8} we simply extrapolate as $N^{-1/\nu d}$ where $\nu d$ is obtained
from Binder's cumulant. Here we cannot do this so must assume a value for
$\nu d$, do the extrapolation and check that the value of $d$ obtained is
consistent. As $c=1$ for both the $Q=4$ Potts and XY models, we use the
value of $\nu d = 2$ reported for the former in \cite{5}.
Then extrapolating $d$ versus $N^{-1/2}$ yields intercept $2.66(3)$,
very close to $d=2.7$ which is usually obtained \cite{5,8,23,24}.
However with $\nu = 1/2$, which is what we have assumed throughout this
analysis, $\nu d = 1.33$ rather than $2$.
Therefore we repeat the fit with $\nu d = 1.33$ and obtain $d = 2.60(3)$.
This is now consistent and we report our best estimate of the internal
fractal dimension for the XY model coupled to 2d quantum gravity as
$d = 2.6(1)$.

\section{Conclusions}

Our numerical results from simulations of the XY model on dynamical
random phi-cubed graphs are consistent with
the persistence of a KT type phase transition when the XY
model is coupled to two-dimensional quantum gravity, in
agreement with the theoretical predictions of \cite{17}.
We found that the critical point moved to lower temperature (from
$\beta_c=2.3(1)$ to $2.4(1)$),
and the peak in the specific heat appeared to be growing more than on a fixed
graph,
but that the latter is probably due to modest statistics.
The simulation on a single fixed graph generated in
a pure two dimensional quantum gravity simulation, as for the Ising
model \cite{19}, gives results consistent with a regular lattice even though
the
status of the continuum limit in this case is uncertain.
The universality in the graph properties we
discovered in our earlier work on various Potts models with
the same value of the central charge does not appear to carry
over to the XY model, as it does not match the other $c=1$ models we have
simulated.

A possible extension of this work would be to couple {\it multiple}
XY models to quantum gravity and see what happens as $c$ increases
beyond $1$. Similar investigations using multiple Potts models
found no dramatic change \cite{8}, but as the XY model does
not appear to share the universal graph properties we found for the Potts
models
its behavior for $c>1$ may be different.
It would also be interesting to use the methods outlined in
this paper to investigate the Villain form of the XY model,
and higher dimensional $O(N)$ spin models, coupled to quantum gravity.

\bigskip
\bigskip
\centerline{\bf Acknowledgements}
\bigskip

This work was supported in part by NATO collaborative research grant CRG910091.
CFB is supported by DOE under contract DE-AC02-86ER40253 and by AFOSR Grant
AFOSR-89-0422.

\vfill
\eject

\bigskip
\bigskip
\centerline{\bf Figure Captions}
\begin{description}

\item[Fig. 1.]
$C$ for fixed simulations; points are from fluctuations in energy,
line is from numerical differentiation of energy for $N=1000$ only.
\item[Fig. 2.]
$[ln(\chi) - ln(C)]/ln(N)] = {(2-\eta) / d}$ versus $\beta$ for
fixed simulations.
\item[Fig. 3.]
$[ln(\chi) - ln(C)]/ln(N)] = {(2-\eta) / d}$ versus $\beta$ for
dynamical simulations.
\item[Fig. 4.]
$C$ for $N=100,500,1000$ dynamical simulations; points are from fluctuations in
energy,
lines from numerical differentiation of energy.
\item[Fig. 5.]
AF, AL and PR3 for $N=2000$ simulation;
the y-scale applies to AF only, AL and PR3 have been scaled
appropriately to fit on plot; $\beta_c$ is indicated by vertical line.
\item[Fig. 6.]
PR3 for XY model along with $Q=2,3,4,10$ Potts models (from \cite{5})
for comparison; lines mark the $\beta_c$'s.

\end{description}
\end{document}